\documentclass{nature}
\usepackage[utf8]{inputenc}
\usepackage[T1]{fontenc}
\usepackage{graphicx}
\usepackage{xcolor}
\usepackage{physics}
\usepackage{bm}
\usepackage{float}
\usepackage{chemformula}

\usepackage{amsfonts}
\usepackage{soul}
\usepackage{hyperref}% add hypertext capabilities
\hypersetup{
    colorlinks=true,
    linkcolor=blue,
    filecolor=blue,
    urlcolor=blue,
   citecolor=blue,
}

\usepackage{upgreek}

\makeatletter
\let\saved@includegraphics\includegraphics
\AtBeginDocument{\let\includegraphics\saved@includegraphics}
\renewenvironment*{figure}{\@float{figure}}{\end@float}
\makeatother

\title{Chiral polariton transport enabled by optical spin Hall effect in perovskite waveguides}

\author{Mateusz Kędziora$^{1}$, Andrzej Opala$^{1,2}$, Maciej Zaremba$^{1}$, Helgi Sigur{\dh}sson$^{1*}$ and Barbara Piętka$^{1*}$}
\begin{document}
\begin{spacing}{1.125}

\maketitle

\begin{scriptsize}
\begin{affiliations}
\item Institute of Experimental Physics, Faculty of Physics, University of Warsaw, ul. Pasteura 5, PL-02-093 Warsaw, Poland
\item Institute of Physics, Polish Academy of Sciences, Aleja Lotników 32/46, PL-02-668 Warsaw, Poland
\end{affiliations}
\end{scriptsize}

\begin{abstract}

Controlling the spin degree of freedom of light at the microscale is crucial for advancing photonic information processing. Spin-polarized light propagation, combined with strong optical nonlinearities, unlocks new functionalities in compact photonic circuits and active spin-optronic devices. Lead halide perovskite exciton–polaritons uniquely combine room-temperature operation, pronounced nonlinearities, and versatile microstructuring, making them a powerful platform for spin-based photonic technologies. Here, we demonstrate polarized edge emission from polariton condensates in perovskite single crystals predesigned into a microwire, forming natural, DBR-free cavity. Above threshold, we observe a distinct waveguiding optical spin Hall effect pattern in both real- and reciprocal-space emission, accompanied by pseudospin phase-locking arising from coherence between opposite edges. Beyond static polarization textures, we achieve spin-resolved polariton edge lasing with chirality exceeding 80\% and spin-polarized signal propagation over tens of micrometres. These results establish CsPbBr$_3$ waveguides as a promising easy-to-fabricate platform for on-chip spin-coded information transport and nonlinear spin-optoelectronics.

\end{abstract}

\maketitle

\section{Introduction}

The development of materials that sustain spin-dependent phenomena in the nonlinear regime is a key driver of progress in spin-optoelectronics. Recent efforts have focused on realizing optical spin Hall and valley Hall effects \cite{Liang2024,Rosiek2023-lb,Peng2024-nu}, generating pure spin-polarized light through circularly polarized (CPL) emission, and implementing spin-selective topological photonic states. Such functionalities are typically achieved by engineering periodic potentials, either through nanostructuring of the substrate\cite{Long2022-wq} or direct patterning of the light-emitting medium\cite{Chen2023-zp}.

A promising platform for realizing spin-polarized states with large optical nonlinear responses is provided by exciton-polaritons (henceforth polaritons), which are the quantum superposition of excitons and photons in confined systems such as microcavities~\cite{Leyder2007-kj}, waveguides~\cite{Walker2019-pp} and metasurfaces~\cite{Dang2022-px}. They exhibit pronounced optical nonlinearities and susceptability towards electric and magnetic fields coming from their exciton component, and rapid timescales and small effective mass from their photonic component~\cite{Carusotto_RMP2013}. They can undergo a power-driven non-equilibrium phase transition into a macroscopic coherent state referred to as a polariton condensate at elevated temperatures that emits coherent light~\cite{Su_NatPhys2020}. In confined photonic systems, such as planar cavities formed from distributed Bragg reflectors, the optical spin Hall effect (OSHE) \cite{Kavokin_PRL2005, Leyder2007-kj, Shi2024} emerges as a key phenomenon in polariton spin transport~\cite{Anton_PRB2015, Sich_ACSPho2018} and topological photonics~\cite{Solnyshkov_OME2021}. The OSHE for polaritons arises from splitting between the in-plane transverse-electric and transverse-magnetic (TE-TM) photon modes, resulting in effective spin-orbit coupling (SOC) of polaritons and separation of coherent spin currents tunable through optical means. OSHE has been demonstrated in GaAs-based microcavities~\cite{Leyder2007-kj}, organic crystals~\cite{Ren2025}, liquid crystal cavities~\cite{Lekenta_LSA2018, Liang2024}, organic microbelts~\cite{Ren_arxiv2024}, and recently in perovskite-filled dielectric cavities~\cite{Shi2024}, to name a few. 
% The light emission is classified as horizontal (H) and vertical (V), respectively, corresponding to TM and TE modes.

Owing to their spatial design flexibility, facile synthesis, scalability, and excellent lasing performance \cite{Deschler2014}, lead–halide perovskites are well established as one of the most important semiconductor platforms in modern photonics, enabling strong polariton nonlinearities and long range transport at room temperature \cite{Su_NatMat2021}. Demonstrating spin phenomena such as the OSHE or circularly polarized light (CPL) emission in perovskites has required substantial effort, including precise crystal or layer microstructuring~\cite{Tian2022-rv}, fabrication of metasurfaces followed by spin-coating of perovskite layers~\cite{Long2022-wq} and adjusting perovskite-based microcavities including incorporation of birefringent materials to tailor the polarization state of the emitted light~\cite{Li2022}. Perovskite microcavities have already proven effective polarization control \cite{Keijsers2025-kx} and linearly polarized lasing \cite{Spencer2021}, owing to their intrinsic in-plane optical birefringence, and have even enabled the realization of a spin-NOT logic gate \cite{Shi2024}. These advances position perovskites as a highly versatile platform for integrating spin-resolved photonic functionalities with nonlinear polariton physics under ambient conditions.

\begin{figure}[H]
\includegraphics[width=0.99\linewidth]{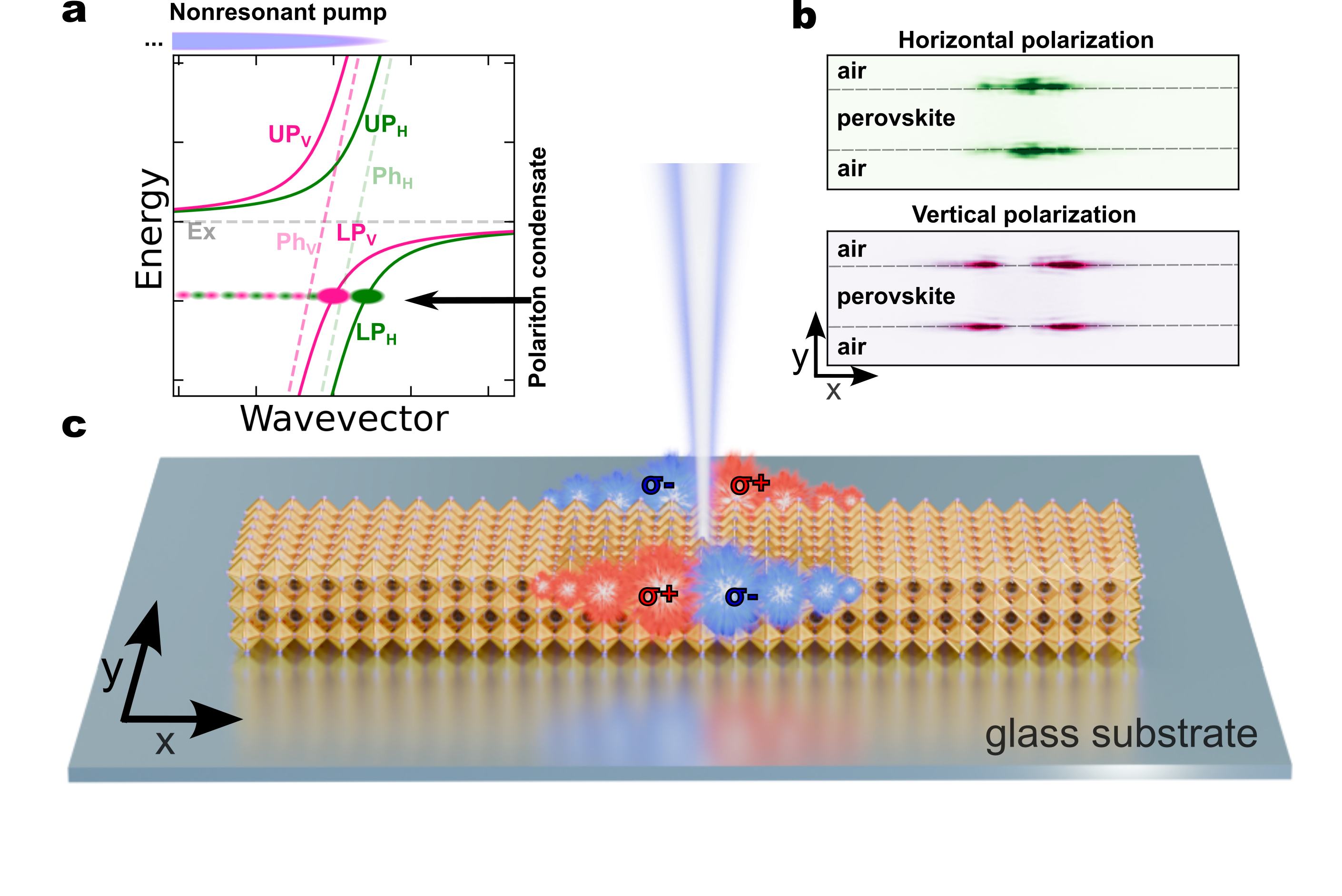}
\label{fig_1}
\caption{\textbf{OSHE in perovskite microwires with TE-TM splitting of waveguide modes.}   
\textbf{a}, Dispersion relation of perovskite microwire with marked H and V polarisations.
\textbf{b}, Real space images of polariton condensate for two linear polarizations.
\textbf{c}, Schematic of the OSHE in perovskite microwire.
}
\end{figure}

In this work, we demonstrate polarized polariton condensate emission at room temperature in the inorganic perovskite crystal CsPbBr$_3$ in the form of rigid waveguides without the need for additional structurization or birefringent dopants. Remarkably, due to the high quality crystal facets, which form a natural optical microcavity, the condensation occurs without external Bragg mirrors. We show simultaneous polariton condensation on two modes, degenerate in energy but with opposite parity and orthogonal polarizations, shown schematically in Fig.~\ref{fig_1}a. In our system, the condensation occurs outside the light cone for the reflected modes, however, edge lasing enables the emission of photons from the polariton condensate, producing a characteristic linear polarization pattern (Fig.~\ref{fig_1}b), accompanied by a pronounced OSHE pattern that emerges after the condensation threshold shown schematically in Fig.~\ref{fig_1}c. This represents a distinct form of OSHE, arising from nonequilibrium condensation of polaritons in the waveguide geometry.  It is worth mentioning that conventional OSHE is reduced in a planar cavity for CsPbBr$_3$ due to the large splitting of the H and V modes at $k=0$~\cite{Shi2024}. Beyond static polarization textures, we further demonstrate that asymmetric nonresonant pumping can drive almost pure chiral condensation (above 80\%) with opposite circular polarizations at the two waveguide edges.

\section{Results}
\begin{figure}[hbt!]
\includegraphics[width=0.99\linewidth]{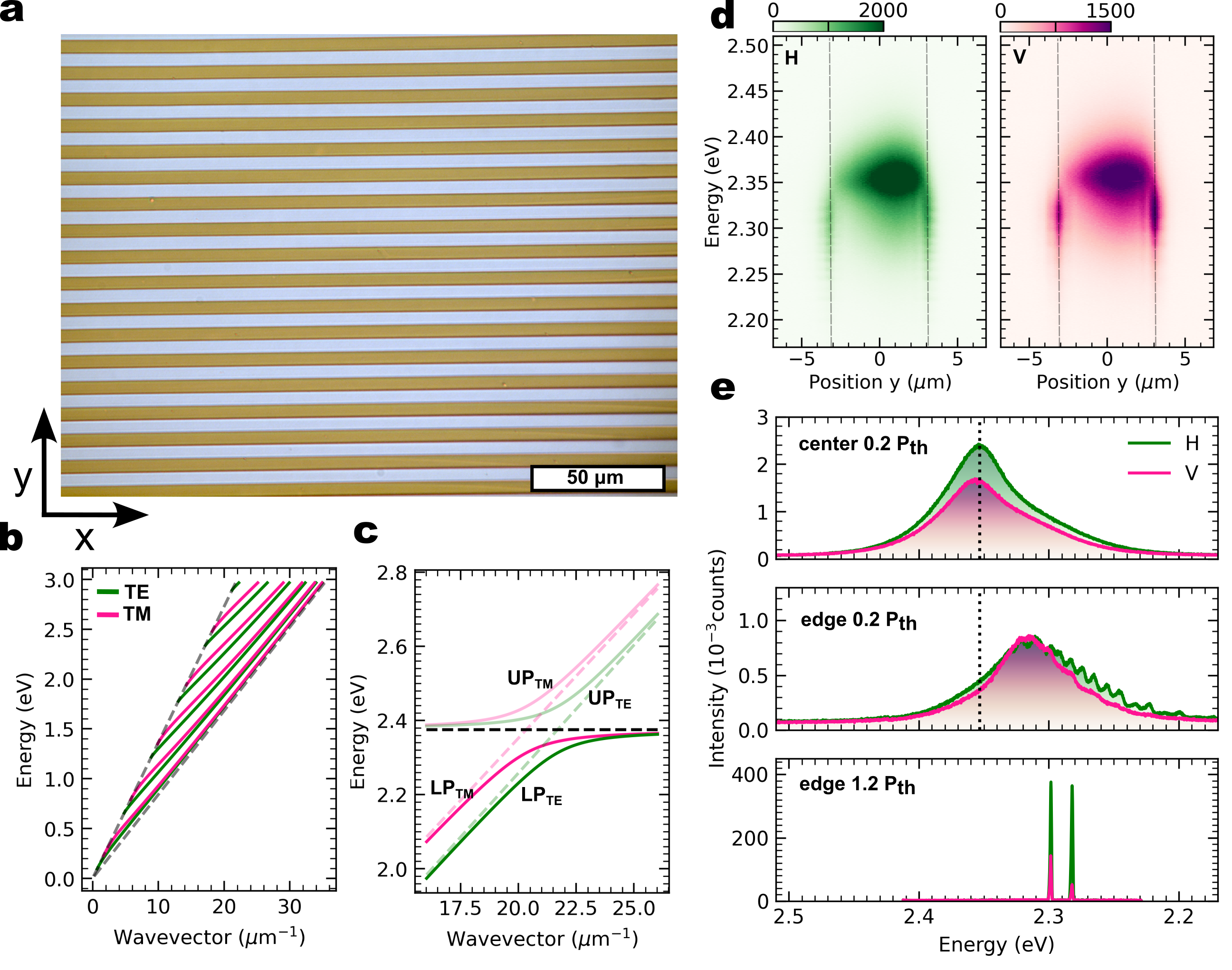}
\caption{\textbf{Polarized condensation in perovskite microwires.}
\textbf{a}, Optical image of a representative CsPbBr\textsubscript{3} crystals with annotated axes.
\textbf{b}, Guided modes of the perovskite slab calculated for the case without exciton,
\textbf{c}, Polariton dispersion calculated for the TE and TM modes inside perovskite. corresponding Hopfields coefficients. 
\textbf{d} real-space PL spectra below the condensation threshold for the two linear polarizations.
\textbf{e} Energy-resolved excitonic (top) and polaritonic (middle) emission at 0.2  $P_\text{th}$ and edge emission 1.2 $P_\text{th}$ (bottom).
}
\label{fig_2}
\end{figure}

We used monocrystalline microwires made of cesium lead bromide to demonstrate polarization patterns in the edge-leaking condensate. The crystals were grown using microfluidically-assisted pseudomorphosis~\cite{Kedziora2024}. Despite the small Q factor (in the range of few hundreds) of such a natural microcavity formed between two opposite edges of the crystal, exciton polaritons are formed even without the use of external DBRs, thanks to the very large Rabi splitting in perovskites~\cite{Kedziora2024,Polimeno2024}. A photo taken with a Nomarski-contrast microscope shows an example crystal in Figure~\ref{fig_2}a. Crystallized from solution at room temperature, \ch{CsPbBr3} has an orthorhombic crystallographic structure in which the perpendicular axes have different lengths, resulting in birefringence of the material with ordinary and extraordinary refractive indices~\cite{Spencer2021,Su2021-qo}. Elongated perovskite crystal wires grown in the order of micrometers in length act as polariton ridge waveguides, but since the guided modes are totally internally reflected, they cannot be probed efficiently through standard far-field spectroscopy measurements. However, dispersion can be seen using a grating on the surface of a waveguide to allow light to couple in and out of the waveguide through higher diffractions orders (Bragg scattering)~\cite{Polimeno2024}. 

Figure~\ref{fig_2}b shows the calculated guided photonic TE and TM modes in a slab perovskite waveguide [see SI for details of solving Maxwell's equations] of thickness $d = 600$ nm (height), width $t = 6$ $\mu$m, deposited on a glass ($n_s = 1.45$) substrate surrounded by air. The refractive index of the perovskite waveguide is $n_c = 2.35$ which correspond to CsPbBr$_3$ operating around 560 nm, sufficiently far away from excitonic resonance~\cite{Ermolaev2023}. Because of the large contrast in the wire's dimensions we can treat its rectangular guided modes as {\it quasi}-TE and TM modes instead (see SI). 

To construct the polariton dispersion relation we assume that we are working sufficiently far from the cut-off frequency of the guided TE and TM modes, and approximate their dispersion as linear around a central wavenumber $\beta_0$, and following the usual semiclassical treatment of strong light-matter coupling, we employ a coupled-oscillator model describing coupling of TE,TM guided photons to corresponding excitons polarized across and along the perovskite microwire. We focus here on the pair of TE and TM of modes which cross the exciton line first.

\begin{equation} \label{eq.H}
    \hat{H} = \begin{pmatrix}
       E_{0}^\text{TE} + v^\text{TE} (\beta - \beta_0) - i \gamma & \Omega & 0 & 0 \\
        \Omega & E_X - i \gamma_X & 0 & 0 \\
        0 & 0 &E_{0}^\text{TM} + v^\text{TM} (\beta - \beta_0) - i\gamma& \Omega \\
        0 & 0 & \Omega & E_X - i \gamma_X
    \end{pmatrix}.
\end{equation}
Here, $v$ and $ E_{0}$ are respectively the group velocity and ground energy of guided photons, $E_X = 2.375$ eV is the exciton energy, $ \Omega = 125$ meV is the Rabi splitting~\cite{Kedziora2024}, and $\gamma$ and $\gamma_X$ are the inverse lifetimes of guided photons and excitons, respectively. We point out that we are not dealing with waveguides of embedded III-V semiconductor quantum wells where excitons are weakly coupled to the TE mode~\cite{Walker2017}, but instead bulk perovskite excitons which should experience similar coupling to both TE and TM modes. Diagonalizing Eq.~\eqref{eq.H} the resulting hybrid light-matter modes are upper (UP) and lower (LP) exciton polariton states corresponding to the eigenstates of the Hamiltonian. The dispersion relation of the polaritons is written,

\begin{equation}\label{eq.pol}
\begin{split}
    E_{\text{UP}}^\alpha & = \frac{E^\alpha + E_X - i (\gamma + \gamma_X)}{2} + \frac{1}{2} \sqrt{ 4 \Omega^2 + [E_X - E^\alpha - i(\gamma - \gamma_X)]^2},\\
    E_{\text{LP}}^\alpha & = \frac{E^\alpha + E_X- i (\gamma + \gamma_X)}{2} - \frac{1}{2} \sqrt{ 4 \Omega^2 + [E_X - E^\alpha - i(\gamma_X - \gamma)]^2},
\end{split}
\end{equation}
where $\alpha \in \{\text{TE,TM}\}$.
Figure~\ref{fig_2}c shows the relevant TE (green dashed line) and TM (pink  dashed line) guided photonic modes centered at $\beta_0 = 21$ $\mu$m$^{-1}$ with respect to the exciton line (black dashed line) and upper and lower polaritons coming from Eqs.~\eqref{eq.pol} marked as solid curves with the same colors as photonic modes. Additional information is given in Figure S2 in the SI showing the corresponding excitonic and photonic (Hopfield) parts, group velocity and TE-TM splitting of the lower polariton modes. For low momenta the lower polaritons are more photon-like and adopt the fast group velocity of the photons, whereas at higher momenta they become more exciton-like and slow down due to the much heavier Wannier-Mott exciton mass.  

To probe photoluminescence of the perovskite, we use a nonresonant, 435 nm V-polarized, picosecond pulsed laser with gaussian beam profile of $6~\mu$m diameter. Spatial and polarization resolved spectra across the microwire for a power corresponding to 0.2 threshold power ($P_\text{th}$ is in the range of 200 $\mu$J/cm$^2$) are shown in Figure~\ref{fig_2}e. Here, vertical (V) and horizontal (H) emission refers to polarization across ($y$) and along ($x$) the wire. %Two main areas of emission are visible - at the center of the wire at the exciton site, the main emission is centered around 2.35 eV and is exciton emission, while polariton emission for lower energies is visible at the edges of the crystal.

The emission originates from two distinct regions: excitonic emission at ~2.35 eV from the wire center, and red-shifted polariton emission from the wire edges. Cross sections taken from the wire center (Figure~\ref{fig_2}e, upper panel) and edges (Figure~\ref{fig_2}e, middle panel) show that exciton emission in orthogonal linear polarizations differs only in intensity, with no spectral shift. In contrast, the polariton emission exhibits similar overall intensity for both polarizations, but distinct spectral profiles due to orthogonal confinement (i.e. across the wire), with Fabry-Perot peaks slightly more closely spaced in energy for horizontal polarization corresponding to the higher wavenumber of TE polaritons for a given energy. It is worth noting that for excitation powers far below threshold, the exciton-related emission directly under the excitation spot is higher than the polariton one at the edges. Above the condensation threshold we can observe nonlinear increase of emission intensity only at the edges which is shown in Figure~\ref{fig_2}e for 1.2 $P_\text{th}$ power (bottom panel). 

Above threshold, condensation of polaritons takes place into two standing wave modes across the microwire with coherent emission from the opposite facing facets~\cite{Kedziora2024}. We note that our far-field measurements allow us to probe the light scattered out from the edges and not the whole standing wave within.  This edge-lasing behaviour shows marked polaritonic condensate features with large blueshift as function of pump intensity, a linewidth collapse and slow broadening with power, a nonlinear input-ouput curve, and redistribution of modal intensities due to thermal-assisted scattering [see SI for details]. The macroscopic coherence of the condensate is evidenced in the sharp fringe contrast when interfering the emission between the opposite facets (discussed in more detail in~\cite{Kedziora2024} and below). In our geometry the horizontal polarization (TE) is found to be several times stronger than the vertical (TM) [see Fig.~\ref{fig_2}e, bottom panel] implying that the TE branch possesses lower losses. Interestingly, in both polarizations the two spectrial peaks for different standing waves are nearly identical despite the different curvatures of the TE and TM dispersion relation. The peaks are separated in energy by approximately $\Delta E \approx 20$ meV corresponding to the free spectral range of standing wave modes across the wire. For more photon-like polaritons, the spectral range can be approximated by a linear dispersion relation of slope $v_\text{gr}$ giving $\Delta E \approx \hbar v_\text{gr} \pi/t$ where $t$ is the wire's width. For $\Delta E \approx 20$ meV we obtain $v_\text{gr} = 58$ $\mu$m ps$^{-1}$ in good agreement with group velocities calculated from Eq.~\ref{eq.pol} [see SI]. Since the energy gap to the next pair of TE and TM modes is $>100$ meV we will assume that the polariton condensate is mostly described within a specific TE-TM branch pair as depicted in Fig.~\ref{fig_2}c. 

\begin{figure}[t]
\includegraphics[width=0.99\linewidth]{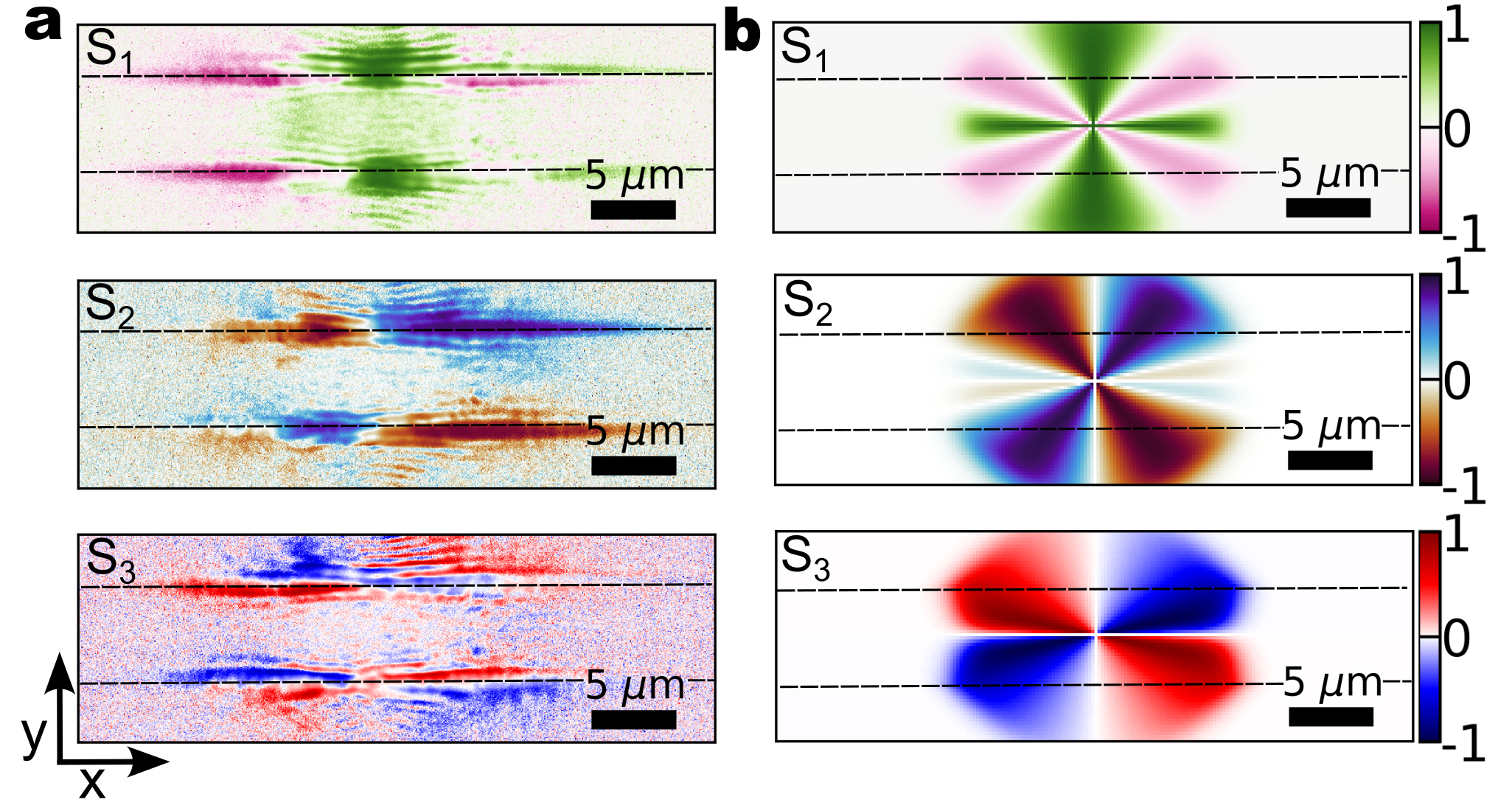}
\caption{\textbf{Optical spin Hall effect in perovskite microwires.}
\textbf{a} Experimental  and \textbf{b} theoretical real-space distributions of the three Stokes parameters evidencing spin-dependent transport of polaritons and resultant anisotropic flower petal polarization emission pattern. The crystals' edges are indicated by dashed lines.}
\label{fig_3}
\end{figure}

We next investigate the spin transport properties of the perovskite microwire polaritons. The polarization degree of freedom of the coherent polariton emission can be directly examined through real-space imaging. The Stokes parameters of light are defined in the standard fashion, 
\begin{equation}
S_{1,2,3}(\mathbf{r}) = \frac{I_{H,D,\sigma^+}(\mathbf{r}) - I_{V,A,\sigma^-}(\mathbf{r})}{I_0(\mathbf{r})}.
\end{equation}
Here, $I_{H,V}, I_{D,A}$, and $I_{\sigma^+,\sigma^-}$ are the intensities of horizontal-vertical, diagonal-antidiagonal, and right-left hand circular polarization components, respectively, normalized against the total intensity $I_0$. Figure~\ref{fig_3}a shows the Stokes parameters of the edge-emitted light measured above the condensation threshold. The results evidence a strongly polarization anisotropic ``flower petal'' emission pattern coming from the wire's edges. The phenomena can be understood in terms of directionally-dependent precession of the polariton condensate macroscopic pseudospin. The pseudospin $\mathbf{S} = \Psi^\dagger \hat{\boldsymbol{\sigma}} \Psi$, where $\hat{\boldsymbol{\sigma}}$ is the Pauli matrix vector, can be in turn defined through a spinor order parameter for the polariton condensate $\Psi(\mathbf{r}) = [\psi_+(\mathbf{r}),\psi_-(\mathbf{r})]^\text{T}$. Namely, $\psi_\pm$ polariton components correspond spin-up and spin-down excitons along the out-of-plane direction of the microwire which are explicitly coupled to $\sigma^\pm$ circularly polarized photons. The Stokes vector therefore gives a direct measure of the polariton condensate spin dynamics through emitted photons. In the circular polarization $\psi_\pm$ basis, the splitting of neighbouring TE-TM polariton modes seen in Fig.~\ref{fig_2}b and~\ref{fig_2}c can be assigned an effective spin-orbit coupling operator~\cite{Kavokin_PRL2005} between $\psi_\pm$ polariton spins giving rise to the optical spin Hall effect (OSHE) initially studied in cryogenic GaAs cavities~\cite{Leyder2007-kj} and recently in room temperature perovskite cavities~\cite{Shi2024},
\begin{equation} \label{eq.OSHE}
    \hat{H}_\text{SOC} = \Delta \begin{pmatrix}
        0 & e^{-2i \varphi} \\
        e^{2 i \varphi} & 0
    \end{pmatrix}.
\end{equation}
Here, $\varphi$ is the azimuthal angle in the $x$-$y$ plane of the waveguide and $2 \Delta$ the amount of TE-TM splitting. In this sense, $\varphi = 0$ corresponds to horizontal and vertically polarized eigenmodes propagating along the $x$-axis. Along the diagonal axis, $\varphi = \pi/4$, the eigenstates are the diagonal and antidiagonal polarized states. The pseudospin dynamics of the condensate can be written in a conservative form~\cite{Shelykh_SST2009} $\partial_t \mathbf{S} = \mathbf{H} \times \mathbf{S}/\hbar$ where $\mathbf{H}$ is the net effective magnetic field felt by polaritons coming from TE-TM splitting and crystal birefringence. The action of $\mathbf{H}$ rotates the spin-state of a propagating polariton until it hits the wire's edge giving rise to the flower pattern in Fig.~\ref{fig_3} (see SI for more details).

In the experiment we used large pump spot similar to the microwire's width. The emission coming from the edge facets across the wire at the location of the pump spot is horizontally polarized. However, the circular polarization detected further away from the pump spot evidences that a portion of the condensate occupies a TM mode, since $\sigma^\pm$ light can only be generated through a superposition of TE and TM light. Assuming a horizontally polarized initial condition at the origin and averaging over rapidly oscillating spatial terms belonging to the standing wave, the condensate ansatz can be approximated in the TE-TM basis as follows,
\begin{equation} \label{eq.cond}
    \Psi = \left[A \cos{(\varphi)} |\text{TM} \rangle + i B \sin{(\varphi)} |\text{TE} \rangle\right] \Theta(\mathbf{r}),
\end{equation}
where $\Theta(\mathbf{r})$ is the slowly varying condensate envelope. It describes the attenuation of guided polaritons along the wire ($x$-direction) and their evanescent field leaking out of the edge $y$-facets. The choice of $\Theta(\mathbf{r})$ does not qualitatively change the results. Since the condensate is mostly populating the TE mode we can set $A=1$ and $B<1$ without loss of generality. In order to calculate the polarization (Stokes vector $\mathbf{S}$) corresponding to our wavefunction~\eqref{eq.cond} we must rotate $\Psi$ into the circular polarization (spin-up spin-down) basis,
\begin{equation} \label{eq.rot}
    \begin{pmatrix}
        \psi_+ \\ 
        \psi_-
    \end{pmatrix} = \frac{1}{\sqrt{2}}\begin{pmatrix}
        e^{i\varphi} & -i e^{i\varphi} \\
        e^{-i\varphi} &  ie^{-i\varphi}
    \end{pmatrix} 
    \begin{pmatrix}
        \psi_\text{TM} \\
        \psi_\text{TE}
    \end{pmatrix}.
\end{equation}
The results of the real space polarization patterns for our wavefunction $\Psi$ are shown in Fig.~\ref{fig_3}b. Notice the polarization around the top and bottom edge of each panel accurately represents the far-field polarization of the light emitted from the wire's edges in experiment. However, our ansatz does not describe reflection of the electromagnetic field with the substrate that can flip $\sigma^\pm \to \sigma^\mp$ and result in reversal of the $S_3$ further away from the wire edges as seen in Fig.~\ref{fig_3}a. To account for this effect more rigorous calculations of Maxwell equations are needed subject to future work. 

\begin{figure}[t]
\centering
\includegraphics[width=0.99\linewidth]{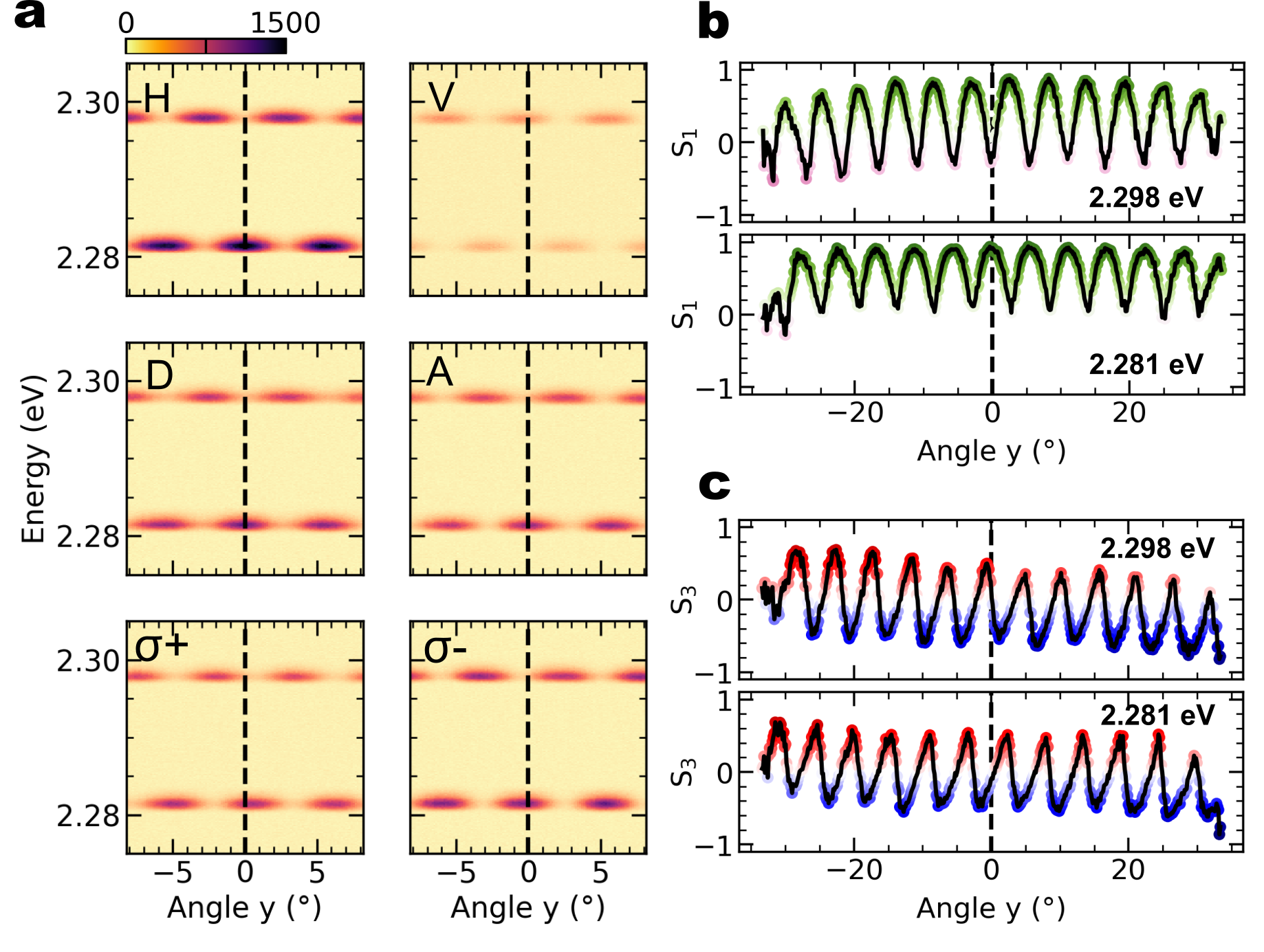}
\caption{\textbf{Pseudospin phase-locking.} 
\textbf{a}, Angle-resolved spectra of the six main polarization components for the two polariton modes in a perovskite microwire.  
\textbf{b}, Stokes parameter $S_1$ for the two modes, with energies indicated in the corresponding insets.  
\textbf{c}, Stokes parameter $S_3$ for the same modes. The vertical black dashed lines at normal incidence are for better visibility.}
\label{fig_4}
\end{figure}

To demonstrate coherence in the polariton emission across the microwire, we angularly resolved the emission spectra. Interference between coherent emission from the two edges of the microwire gives rise to a distinct fringe pattern in the far field (reciprocal space), with the fringe spacing determined by the crystal wire width. The clear contrast between the fringes evidences that light emitted from the two edges is coherent and phase-locked. We note that wider microwires can enable multi-wavelength emission (condensation into a ladder of modes) which blurs the intensity of OSHE (see SI for the details), as well as wavelength switching driven by mode competition~\cite{Kedziora2024}.

Figure~\ref{fig_4}a shows the spectra measured at a pump power of  1.2 P$_{th}$, resolved into its six principal polarizations, for a relatively narrow perovskite microwire of width 6 $\mu$m within a small range of angles for greater clarity. Figures~\ref{fig_4}b,c show the corresponding $S_1$ and $S_3$ Stokes parameters in a wider range of angles. As discussed in Fig.~\ref{fig_2}e, the condensate can be divided between two neighbouring standing waves across the wire resulting in two spectral peaks seen as two lines in Fig.~\ref{fig_4}a. An odd parity standing wave corresponds to a far field emission with minimum intensity at normal incidence (emission from wire edges is anti-phase), and an even parity standing wave as one with a maximum intensity at normal incidence (emission from wire edges is in-phase). The measurements in Figure~\ref{fig_4}a reveal that the energy-degenerate H and V modes ($S_1$) exhibit opposite parity: at 2.298 eV the horizontal mode is odd while the vertical mode is even, whereas for the next lasing mode at 2.281 eV the behavior is reversed. For diagonal polarizations ($S_2$), the parity of the modes is the same, while for the circular polarization components ($S_3$) no clear even or odd character is observed, as the intensity at $\theta = 0$ shows neither a maximum nor a minimum. This asymmetric emission profile in the far field circular polarization is a consequence that circular polarization is a superposition of linearly polarized light $|\sigma^\pm\rangle = \frac{1}{\sqrt{2}}(|H\rangle \mp i |V\rangle)$ and therefore both even- and odd-parity mode profiles in $|H\rangle$ and $|V\rangle$ contribute to the far field circular emission profile.

To further probe the coherence of the emission, we measured the full reciprocal-space emission, which, due to the sample structure, shows a strong dependence on the angular distribution of the emitted light.
Figure S7 in SI shows the Stokes parameters for a 6 $\mu$m wide wire and a pumping power of about 3.6 P$_{th}$ corresponding to results in Figure 4. 
 
\begin{figure}[H]
\includegraphics[width=0.99\linewidth]{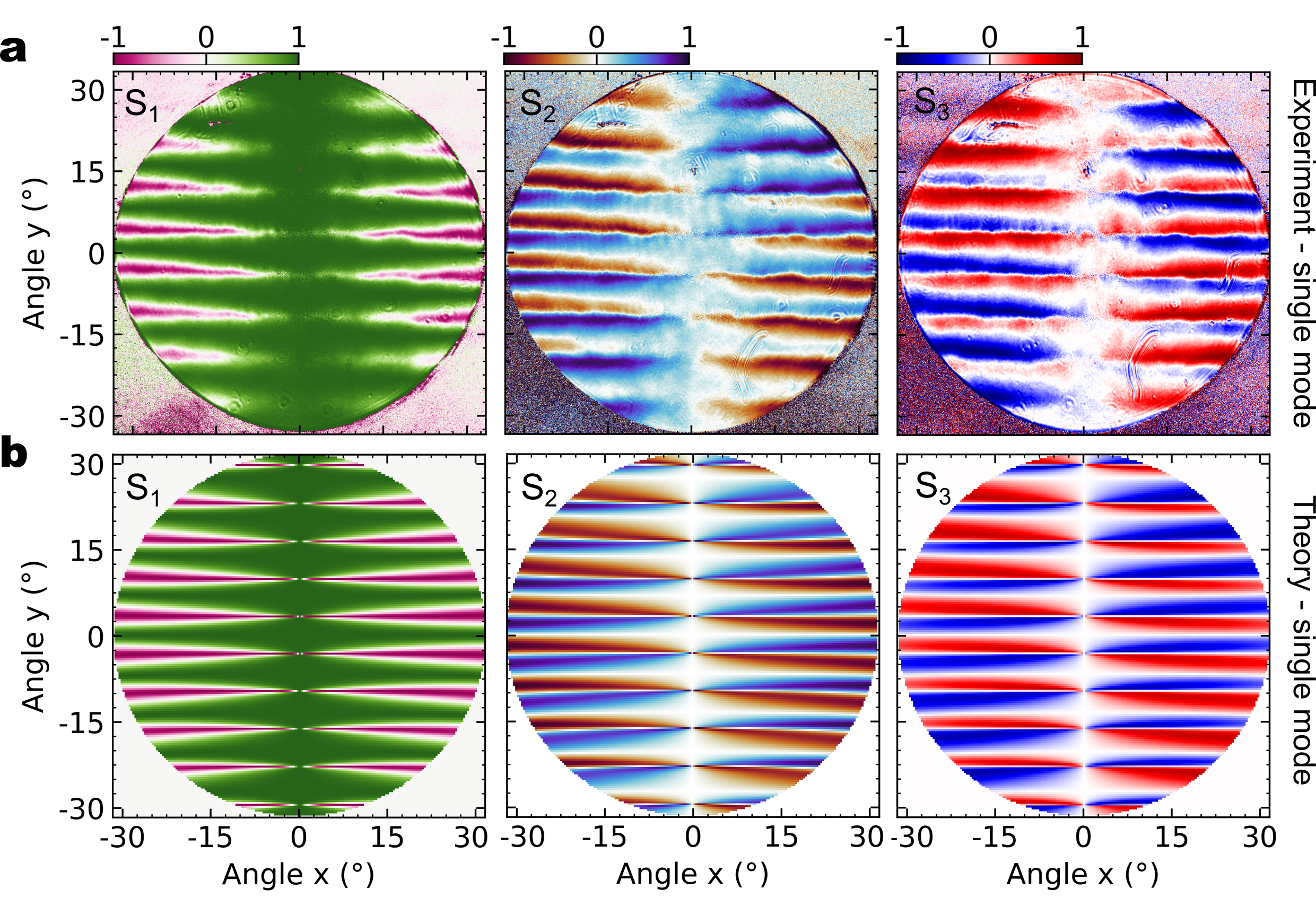}
\caption{\textbf{Reciprocal-space polarization patterns of coherent polariton edge emission.}
 Experimental distributions of the three Stokes parameters $S_{1,2,3}$ for a polariton microwire pumped above condensation threshold. The wire width is $t \approx 5$ $\mu m$. The appearance of fringes in the polarization profile implies coherent (phase-locked) emission from the wire edges which superimposes into a beating polarization pattern determined by the OSHE and wire width. 
\textbf{b} Corresponding calculated Stokes far field pattern by filtering Eq.~\eqref{eq.cond} around the wire edges and taking the Fourier transform.}
\label{fig_5}
\end{figure}

However, the presence of two modes blurs the far field pattern in time-integrated measurements. To overcome this we chose a slightly narrower wire of width $t = 5.0$ $\mu$m such that neighboring standing waves are spectrally separated more and we obtain clear single-mode emission~\cite{Kedziora2024}. This results in clearer polarization pattern in the far field, shown in Figure~\ref{fig_5}a for all Stokes components. At small angles along the wire $\theta_x \approx 0$ the H-polarization is dominant, corresponding to strong condensate emission from a TE polarized standing wave. At higher angles $|\theta_x|>0$ polaritons undergo spin-precession within their lifetime resulting in V-polarized emission as fringes with spacing $\Delta k = \pi /t = 0.63$ $\mu$m$^{-1}$. Importantly, we observe the momentum-space separation of the polariton spins seen both in the diagonal ($S_2$) and circular ($S_3$) Stokes components forming the characteristic quadrupole typical of the OSHE in planar cavities~\cite{Leyder2007-kj}.  
% The behaviour of circular polarisations ($S_3$), changing their sign for transitions along angle $\theta_x=0$, confirms spin-orbit coupling. 
The far field emission pattern can be theoretically reconstructed by performing Gaussian filtering of our real space ansatz [Eq.~\eqref{eq.cond}] around $y = \pm 2.5$ $\mu$m and taking the Fourier transform. The results are presented in Fig.~\ref{fig_5}c showing good agreement with experiment.

\begin{figure}[H]
\includegraphics[width=0.99\linewidth]{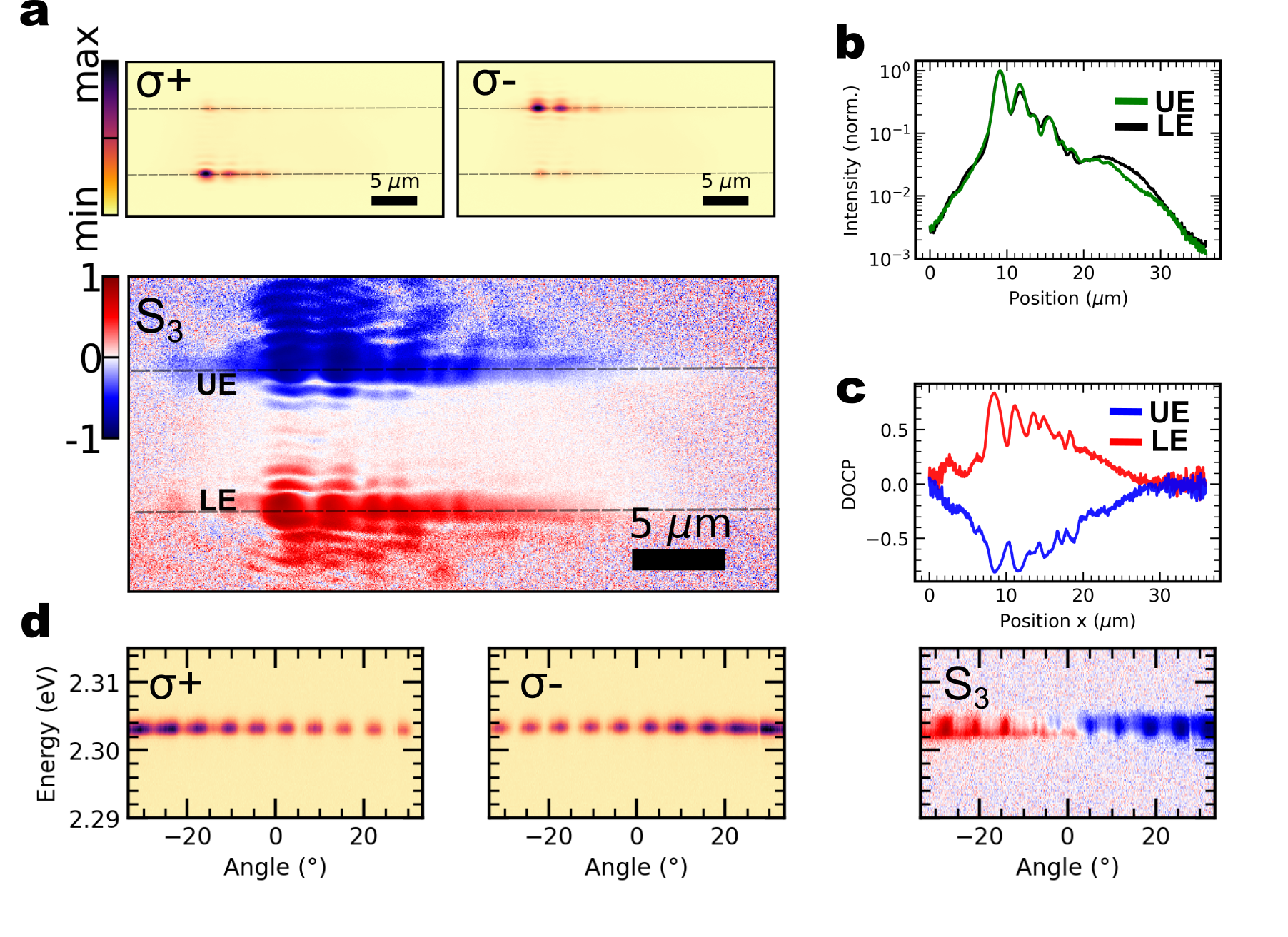}
\caption{\textbf{Chiral polariton transport induced by asymmetrical pump on OSHE dominated modes in perovskite waveguide.}  
\textbf{a} Measured polariton condensate resolved into left- and right-circular polarization components (top) with corresponding Stokes parameter $S_3$ (bottom). Cross-sections of normalized intensity  
\textbf{b}, and $S_3$ \textbf{c} at the two lateral edges of the microstructure labeled UE and LE.  
\textbf{d} Angle-resolved measurements in circular polarizations (top) and the corresponding $S_3$ parameter (bottom).}
\label{fig_6}
\end{figure}

The OSHE effect, in combination with the structure of CsPbBr$_3$ microwires, allows us to demonstrate spin-separated polariton transport in a waveguide geometry shown in Fig.~\ref{fig_6}.
By employing asymmetric optical pumping, we realize an edge splitter that selectively favors polariton propagation in one direction which separate into opposite spin currents, due to the OSHE, observable along the two waveguide edges. 
To ensure spatially homogeneous signal collection, asymmetric pumping was performed in a transmission configuration using Gaussian beam elongated in $y$-direction and excitation at a small (less than 5$^{\circ}$) angle. The circularly polarised emission distributions above the condensation threshold are shown in Fig.~\ref{fig_6}a. The intensity cross-section taken along the uppper (UE) and lower (LE) edges (see Fig.~\ref{fig_6}b), shows a gradual decay over 20~$\mu m$ and intensity beats due to the interference between the wire walls. While the total intensity is nearly identical on the two edges, the $S_3$ parameter (Fig.~\ref{fig_6}c) has equal magnitude but opposite sign, reaching $|S_3| = 0.85$, among the highest values reported for room-temperature polariton systems. Angle-resolved spectra in Fig.~\ref{fig_6}d show a pronounced asymmetry in the distribution of the two circular polarizations, in sharp contrast to the more balanced emission observed under symmetric pumping conditions illustrated in Fig.~\ref{fig_4}. We note that our experiments are performed over multiple realizations of the condensate (multiple pulses) and, due to spontaneous symmetry breaking upon condensation, one might expect equal amounts of polaritons, on average, flowing left and right. Under this condition one retrieves the four-lobed $S_3$ pattern as seen in Fig.~\ref{fig_3}. However, the clear $k_x>0$ (momentum along wire to the right) preference of the emission underlines the bias from the pump on the condensate. The scale of our separated spin-currents is similar to what has been reported in planar perovskite cavities~\cite{Shi2024, Liang2024} but a clear advantage of our platform is that coherence emerges spontaneously (i.e., the laser imprints no information) with clear nonlinear behaviour as function of nonresonant pump power around condensation, as well as requires no vertical stacks as mirrors and instead strong-coupling forms under total-internal reflection in the waveguide~\cite{Kedziora2024}.

\section{Conclusions}

In summary, we have demonstrated a pronounced OSHE in polariton condensates formed in CsPbBr$_3$ crystals shaped into waveguides and operating at room temperature. By generating an extended polariton condensate, probed through edge lasing and polarization-resolved measurements, we revealed spin textures characteristic of the OSHE in both real and reciprocal space. These experimental observations are supported by theoretical calculations of polarization patterns for waveguide polaritons on TE and TM modes.

Unlike earlier realizations, primarily in GaAs-based planar microcavities at cryogenic temperatures, our approach requires neither distributed Bragg reflectors nor complex heterostructures, enabling a far simpler and more scalable platform. The use of non-resonant pumping and non-equilibrium condensation places the effect firmly in the nonlinear regime, opening new opportunities for nonlinear spin–optronic functionalities. Taking profit on this capability and waveguide pre-designed shape, we demonstrated edge-separated circularly polarized condensates with high chirality and long-range propagation. This combination of room-temperature operation, strong nonlinearity, and spin-selective transport paves the way for practical spin filters and spin splitters for future spin-based photonic and computing architectures.

\section*{Methods}
\textit{Crystallization of perovskites}\\
Perovskite micro-wires were produced by crystallization in spatially defined templates made of polydimethylsiloxane (PDMS). Uncured PDMS (Sylgard 184,9:1 base to curring agent) was placed on a gallium arsenide master template and heated for 10 minutes at 140 degrees Celsius to solidify. The finished template was then transferred to glass slide that had been washed successively in soapy water, acetone, and isopropanol. The perovskite solution was prepared by mixing equal amounts of CsBr and \ch{PbBr2} (both from Alfa Aesar) in DMSO (Sigma Aldrich). Approximately 1 microliter of a 0.2 molar \ch{CsPbBr3} solution was added to the template, which was then sealed in a plastic container and stored in a refrigerator for 1 day, followed by 2 days at room temperature. After opening, slow crystallization occurs over approximately 1-2 days.\\
\vspace{1mm}

\noindent \textit{Optical measurements}\\
The excitation laser was a Fluence Jasper X0 with a Fluence Harmony optical parametric amplifier and collecting objective Nikon LU Plan ELWD 50$\times$ magnification objective with a numerical aperture of 0.55. The signal was collected on high-resolution Andor Marana CMOS camera or Princeton Ins. SpectraPro HRS500 spectrometer with PIXIS CCD camera. Polarization optics (half waveplate, quater waveplate and linear polarizer) callibrated to resolve six main polarizations were placed at the detection path.

\section*{References}
\vspace{10mm}
\bibliographystyle{naturemag}
\bibliography{_biblio}

\begin{thebibliography}{10}
\expandafter\ifx\csname url\endcsname\relax
  \def\url#1{\texttt{#1}}\fi
\expandafter\ifx\csname urlprefix\endcsname\relax\def\urlprefix{URL }\fi
\providecommand{\bibinfo}[2]{#2}
\providecommand{\eprint}[2][]{\url{#2}}

\bibitem{Liang2024}
\bibinfo{author}{Liang, J.} \emph{et~al.}
\newblock \bibinfo{title}{Polariton spin hall effect in a rashba–dresselhaus regime at room temperature}.
\newblock \emph{\bibinfo{journal}{Nature Photonics}} \textbf{\bibinfo{volume}{18}}, \bibinfo{pages}{357–362} (\bibinfo{year}{2024}).
\newblock \urlprefix\url{http://dx.doi.org/10.1038/s41566-023-01375-x}.

\bibitem{Rosiek2023-lb}
\bibinfo{author}{Rosiek, C.~A.} \emph{et~al.}
\newblock \bibinfo{title}{Observation of strong backscattering in valley-hall photonic topological interface modes}.
\newblock \emph{\bibinfo{journal}{Nat. Photonics}} \textbf{\bibinfo{volume}{17}}, \bibinfo{pages}{386--392} (\bibinfo{year}{2023}).

\bibitem{Peng2024-nu}
\bibinfo{author}{Peng, K.} \emph{et~al.}
\newblock \bibinfo{title}{Topological valley hall polariton condensation}.
\newblock \emph{\bibinfo{journal}{Nat. Nanotechnol.}} \textbf{\bibinfo{volume}{19}}, \bibinfo{pages}{1283--1289} (\bibinfo{year}{2024}).

\bibitem{Long2022-wq}
\bibinfo{author}{Long, G.} \emph{et~al.}
\newblock \bibinfo{title}{Perovskite metasurfaces with large superstructural chirality}.
\newblock \emph{\bibinfo{journal}{Nat. Commun.}} \textbf{\bibinfo{volume}{13}}, \bibinfo{pages}{1551} (\bibinfo{year}{2022}).

\bibitem{Chen2023-zp}
\bibinfo{author}{Chen, Y.} \emph{et~al.}
\newblock \bibinfo{title}{Compact spin-valley-locked perovskite emission}.
\newblock \emph{\bibinfo{journal}{Nat. Mater.}} \textbf{\bibinfo{volume}{22}}, \bibinfo{pages}{1065--1070} (\bibinfo{year}{2023}).

\bibitem{Leyder2007-kj}
\bibinfo{author}{Leyder, C.} \emph{et~al.}
\newblock \bibinfo{title}{Observation of the optical spin hall effect}.
\newblock \emph{\bibinfo{journal}{Nat. Phys.}} \textbf{\bibinfo{volume}{3}}, \bibinfo{pages}{628--631} (\bibinfo{year}{2007}).

\bibitem{Walker2019-pp}
\bibinfo{author}{Walker, P.~M.} \emph{et~al.}
\newblock \bibinfo{title}{Spatiotemporal continuum generation in polariton waveguides}.
\newblock \emph{\bibinfo{journal}{Light Sci. Appl.}} \textbf{\bibinfo{volume}{8}}, \bibinfo{pages}{6} (\bibinfo{year}{2019}).

\bibitem{Dang2022-px}
\bibinfo{author}{Dang, N. H.~M.} \emph{et~al.}
\newblock \bibinfo{title}{Realization of polaritonic topological charge at room temperature using polariton bound states in the continuum from perovskite metasurface}.
\newblock \emph{\bibinfo{journal}{Adv. Opt. Mater.}} \textbf{\bibinfo{volume}{10}}, \bibinfo{pages}{2102386} (\bibinfo{year}{2022}).

\bibitem{Carusotto_RMP2013}
\bibinfo{author}{Carusotto, I.} \& \bibinfo{author}{Ciuti, C.}
\newblock \bibinfo{title}{Quantum fluids of light}.
\newblock \emph{\bibinfo{journal}{Rev. Mod. Phys.}} \textbf{\bibinfo{volume}{85}}, \bibinfo{pages}{299--366} (\bibinfo{year}{2013}).
\newblock \urlprefix\url{https://link.aps.org/doi/10.1103/RevModPhys.85.299}.

\bibitem{Su_NatPhys2020}
\bibinfo{author}{Su, R.} \emph{et~al.}
\newblock \bibinfo{title}{Observation of exciton polariton condensation in a perovskite lattice at room temperature}.
\newblock \emph{\bibinfo{journal}{Nature Physics}} \textbf{\bibinfo{volume}{16}}, \bibinfo{pages}{301--306} (\bibinfo{year}{2020}).
\newblock \urlprefix\url{https://doi.org/10.1038/s41567-019-0764-5}.

\bibitem{Kavokin_PRL2005}
\bibinfo{author}{Kavokin, A.}, \bibinfo{author}{Malpuech, G.} \& \bibinfo{author}{Glazov, M.}
\newblock \bibinfo{title}{Optical spin hall effect}.
\newblock \emph{\bibinfo{journal}{Phys. Rev. Lett.}} \textbf{\bibinfo{volume}{95}}, \bibinfo{pages}{136601} (\bibinfo{year}{2005}).
\newblock \urlprefix\url{https://link.aps.org/doi/10.1103/PhysRevLett.95.136601}.

\bibitem{Shi2024}
\bibinfo{author}{Shi, Y.} \emph{et~al.}
\newblock \bibinfo{title}{Coherent optical spin hall transport for polaritonics at room temperature}.
\newblock \emph{\bibinfo{journal}{Nature Materials}} \textbf{\bibinfo{volume}{24}}, \bibinfo{pages}{56–62} (\bibinfo{year}{2024}).
\newblock \urlprefix\url{http://dx.doi.org/10.1038/s41563-024-02028-2}.

\bibitem{Anton_PRB2015}
\bibinfo{author}{Ant\'on, C.} \emph{et~al.}
\newblock \bibinfo{title}{Optical control of spin textures in quasi-one-dimensional polariton condensates}.
\newblock \emph{\bibinfo{journal}{Phys. Rev. B}} \textbf{\bibinfo{volume}{91}}, \bibinfo{pages}{075305} (\bibinfo{year}{2015}).
\newblock \urlprefix\url{https://link.aps.org/doi/10.1103/PhysRevB.91.075305}.

\bibitem{Sich_ACSPho2018}
\bibinfo{author}{Sich, M.} \emph{et~al.}
\newblock \bibinfo{title}{Spin domains in one-dimensional conservative polariton solitons}.
\newblock \emph{\bibinfo{journal}{ACS Photonics}} \textbf{\bibinfo{volume}{5}}, \bibinfo{pages}{5095–5102} (\bibinfo{year}{2018}).
\newblock \urlprefix\url{http://dx.doi.org/10.1021/acsphotonics.8b01410}.

\bibitem{Solnyshkov_OME2021}
\bibinfo{author}{Solnyshkov, D.~D.} \emph{et~al.}
\newblock \bibinfo{title}{Microcavity polaritons for topological photonics [invited]}.
\newblock \emph{\bibinfo{journal}{Optical Materials Express}} \textbf{\bibinfo{volume}{11}}, \bibinfo{pages}{1119} (\bibinfo{year}{2021}).
\newblock \urlprefix\url{http://dx.doi.org/10.1364/OME.414890}.

\bibitem{Ren2025}
\bibinfo{author}{Ren, J.} \emph{et~al.}
\newblock \bibinfo{title}{Optical spin hall effect pattern switching in polariton condensates in organic single-crystal microbelts}.
\newblock \emph{\bibinfo{journal}{Journal of the American Chemical Society}} \textbf{\bibinfo{volume}{147}}, \bibinfo{pages}{8336–8342} (\bibinfo{year}{2025}).
\newblock \urlprefix\url{http://dx.doi.org/10.1021/jacs.4c15894}.

\bibitem{Lekenta_LSA2018}
\bibinfo{author}{Lekenta, K.} \emph{et~al.}
\newblock \bibinfo{title}{Tunable optical spin hall effect in a liquid crystal microcavity}.
\newblock \emph{\bibinfo{journal}{Light: Science {\&} Applications}} \textbf{\bibinfo{volume}{7}} (\bibinfo{year}{2018}).
\newblock \urlprefix\url{http://dx.doi.org/10.1038/s41377-018-0076-z}.

\bibitem{Ren_arxiv2024}
\bibinfo{author}{Ren, J.} \emph{et~al.}
\newblock \bibinfo{title}{Optical spin hall effect pattern switching in polariton condensates in organic single-crystal microbelts} (\bibinfo{year}{2024}).
\newblock \urlprefix\url{https://arxiv.org/abs/2401.03877}.

\bibitem{Deschler2014}
\bibinfo{author}{Deschler, F.} \emph{et~al.}
\newblock \bibinfo{title}{High photoluminescence efficiency and optically pumped lasing in solution-processed mixed halide perovskite semiconductors}.
\newblock \emph{\bibinfo{journal}{The Journal of Physical Chemistry Letters}} \textbf{\bibinfo{volume}{5}}, \bibinfo{pages}{1421--1426} (\bibinfo{year}{2014}).
\newblock \urlprefix\url{https://doi.org/10.1021/jz5005285}.

\bibitem{Su_NatMat2021}
\bibinfo{author}{Su, R.} \emph{et~al.}
\newblock \bibinfo{title}{Perovskite semiconductors for room-temperature exciton-polaritonics}.
\newblock \emph{\bibinfo{journal}{Nature Materials}} \textbf{\bibinfo{volume}{20}}, \bibinfo{pages}{1315--1324} (\bibinfo{year}{2021}).
\newblock \urlprefix\url{https://doi.org/10.1038/s41563-021-01035-x}.

\bibitem{Tian2022-rv}
\bibinfo{author}{Tian, J.} \emph{et~al.}
\newblock \bibinfo{title}{Optical rashba effect in a light-emitting perovskite metasurface}.
\newblock \emph{\bibinfo{journal}{Adv. Mater.}} \textbf{\bibinfo{volume}{34}}, \bibinfo{pages}{e2109157} (\bibinfo{year}{2022}).

\bibitem{Li2022}
\bibinfo{author}{Li, Y.} \emph{et~al.}
\newblock \bibinfo{title}{Manipulating polariton condensates by {R}ashba-{D}resselhaus coupling at room temperature}.
\newblock \emph{\bibinfo{journal}{Nature Communications}} \textbf{\bibinfo{volume}{13}} (\bibinfo{year}{2022}).
\newblock \urlprefix\url{https://doi.org/10.1038/s41467-022-31529-4}.

\bibitem{Keijsers2025-kx}
\bibinfo{author}{Keijsers, G.} \emph{et~al.}
\newblock \bibinfo{title}{Continuous-wave nonlinear polarization control and signatures of criticality in a perovskite cavity}.
\newblock \emph{\bibinfo{journal}{Nat. Photonics}} \textbf{\bibinfo{volume}{19}}, \bibinfo{pages}{733--739} (\bibinfo{year}{2025}).

\bibitem{Spencer2021}
\bibinfo{author}{Spencer, M.~S.} \emph{et~al.}
\newblock \bibinfo{title}{Spin-orbit–coupled exciton-polariton condensates in lead halide perovskites}.
\newblock \emph{\bibinfo{journal}{Science Advances}} \textbf{\bibinfo{volume}{7}}, \bibinfo{pages}{eabj7667} (\bibinfo{year}{2021}).
\newblock \urlprefix\url{https://www.science.org/doi/abs/10.1126/sciadv.abj7667}.
\newblock \eprint{https://www.science.org/doi/pdf/10.1126/sciadv.abj7667}.

\bibitem{Kedziora2024}
\bibinfo{author}{Kędziora, M.} \emph{et~al.}
\newblock \bibinfo{title}{Predesigned perovskite crystal waveguides for room-temperature exciton–polariton condensation and edge lasing}.
\newblock \emph{\bibinfo{journal}{Nat. Mater.}} \textbf{\bibinfo{volume}{23}}, \bibinfo{pages}{1476--1480} (\bibinfo{year}{2024}).

\bibitem{Polimeno2024}
\bibinfo{author}{Polimeno, L.} \emph{et~al.}
\newblock \bibinfo{title}{Room temperature polariton condensation from whispering gallery modes in cspbbr3 microplatelets}.
\newblock \emph{\bibinfo{journal}{Advanced Materials}} \textbf{\bibinfo{volume}{36}} (\bibinfo{year}{2024}).
\newblock \urlprefix\url{http://dx.doi.org/10.1002/adma.202312131}.

\bibitem{Su2021-qo}
\bibinfo{author}{Su, R.} \emph{et~al.}
\newblock \bibinfo{title}{Direct measurement of a non-hermitian topological invariant in a hybrid light-matter system}.
\newblock \emph{\bibinfo{journal}{Sci. Adv.}} \textbf{\bibinfo{volume}{7}}, \bibinfo{pages}{eabj8905} (\bibinfo{year}{2021}).

\bibitem{Ermolaev2023}
\bibinfo{author}{Ermolaev, G.} \emph{et~al.}
\newblock \bibinfo{title}{Giant and tunable excitonic optical anisotropy in single-crystal halide perovskites}.
\newblock \emph{\bibinfo{journal}{Nano Letters}} \textbf{\bibinfo{volume}{23}}, \bibinfo{pages}{2570–2577} (\bibinfo{year}{2023}).
\newblock \urlprefix\url{http://dx.doi.org/10.1021/acs.nanolett.2c04792}.

\bibitem{Walker2017}
\bibinfo{author}{Walker, P.~M.} \emph{et~al.}
\newblock \bibinfo{title}{Dark solitons in high velocity waveguide polariton fluids}.
\newblock \emph{\bibinfo{journal}{Phys. Rev. Lett.}} \textbf{\bibinfo{volume}{119}}, \bibinfo{pages}{097403} (\bibinfo{year}{2017}).
\newblock \urlprefix\url{https://link.aps.org/doi/10.1103/PhysRevLett.119.097403}.

\bibitem{Shelykh_SST2009}
\bibinfo{author}{Shelykh, I.~A.}, \bibinfo{author}{Kavokin, A.~V.}, \bibinfo{author}{Rubo, Y.~G.}, \bibinfo{author}{Liew, T. C.~H.} \& \bibinfo{author}{Malpuech, G.}
\newblock \bibinfo{title}{Polariton polarization-sensitive phenomena in planar semiconductor microcavities}.
\newblock \emph{\bibinfo{journal}{Semiconductor Science and Technology}} \textbf{\bibinfo{volume}{25}}, \bibinfo{pages}{013001} (\bibinfo{year}{2009}).
\newblock \urlprefix\url{http://dx.doi.org/10.1088/0268-1242/25/1/013001}.

\end{thebibliography}

\section*{Data availability}
\noindent All data that supports the conclusions of this study are included in the article. The data presented in this study are available from the corresponding author upon reasonable request.

\section*{Acknowledgments} 
\noindent We acknowledge Anna Szerling and Marek Ekielski from the Łukasiewicz Research Network – Institute of Microelectronics and Photonics for the technology support. H.S. acknowledges the project No. 2022/45/P/ST3/00467 co-funded by the Polish National Science Center and the European Union Framework Programme for Research and Innovation Horizon 2020 under the Marie Skłodowska-Curie grant agreement No. 945339. M.K. acknowledges the project No. 2022/47/B/ST3/02411 funded by the National Science Center, Poland and  support from the Foundation for Polish Science (FNP). B.P. acknowledges the project No. 2020/37/B/ST3/01657 funded by the National Science Center, Poland. This work was supported by the European Union EIC Pathfinder Open project “Polariton Neuromorphic Accelerator” (PolArt, Id: 101130304). A.O. acknowledges the project No. 2024/52/C/ST3/00324 funded by the National Science Center, Poland and support from the Foundation for Polish Science (FNP).

\section*{Author contributions}
\noindent M.K. and B.P. conceived the idea; M.K. grew perovskite crystals and performed the optical experiments; H.S. developed the theoretical description with input from A.O.; M.Z. participated in optical measurements; M.K., H.S. and A.O. visualised the data; M.K. and H.S. wrote the manuscript with input from other authors; B.P. supervised the project.

\section*{Competing interests} 
\noindent The authors declare no competing interests.

\section*{Additional information} 
\noindent{\bf Correspondence and requests for materials} should be addressed to H.S. and B.P.

\end{spacing}
\end{document}